\documentclass{caosp306}
\usepackage{graphicx}
\usepackage{natbib}
\articleNo{123}
\pubyear{2005}
\volume{35}
\volnumber{3}
\firstpage{1}
\received{May 1, 2005}
\accepted{August 28, 2005}

\def\BibTeX{{\rm B\kern-.05em{\sc i\kern-.025em b}\kern-.08em
             T\kern-.1667em\lower.7ex\hbox{E}\kern-.125emX}}

\begin{document}

\hauthor{Z.\,Mikul\'a\v{s}ek and M.\,Skarka}

\title{How far can we trust published TESS periods?}

\author{
        Zden\v{e}k Mikul\'a\v{s}ek \inst{1}
      \and
        Marek Skarka \inst{1,2}
       }

\institute{
           Department of Theoretical Physics an Astrophysics, Masaryk University, Kotl\'a\v{r}sk\'a 2, 611~38 Brno, The Czech Republic \email{mikulas@physics.muni.cz}\\
         \and
           \ondrejov
          }

\date{March 8, 2003}

\maketitle

\begin{abstract}
Possible inaccuracies in the determination of periods from short-term time series caused by disregard of the real course of light curves and instrumental trends are documented on the example of the period analysis of simulated TESS-like light curve by notorious Lomb-Scargle method.
\keywords{variable stars -- period analysis -- TESS data}
\end{abstract}

\section{Introduction}
\label{intr}
TESS data are now one of the most popular sources of information about variable stars, including their periods. However, TESS data suffers from two shortcomings that significantly corrupt the results of the period analysis with the standard tools. The data are strongly affected by instrumental trends of various kinds and are mostly obtained only in non-standard short time interval of 27 days (with a break) that is often comparable with the lengths of the periods themselves. The majority of such days-long periods and their uncertainties are mere artifacts of the method used to determine them and, at best, are just estimates.

The most widely used instruments of period analysis are frequency periodograms of various kinds, where astrophysically important frequencies are those in which the extremes of some suitably chosen characteristic of the studied data set occur. The reason that they generally give virtually identical results is that they are mostly consciously or covertly based on data interleaving by sine/cosine functions using the least squares method. The most commonly used is the historically oldest Lomb-Scargle method \citep{press89}, which after some algebra can be converted to a more informative version periodogram, where the frequency-dependent characteristic is the amplitude of light or other changes \citep{mikpau15}.

\begin{figure}
\centerline{\includegraphics[width=1\textwidth,clip=]{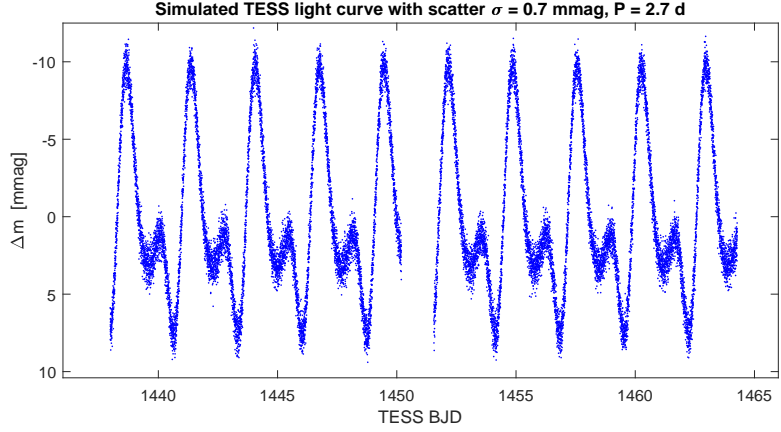}}
\caption{Simulated light curve without trends represented by 17\,925 points. }
\label{LCshow}
\end{figure}

The following demonstration, based on the simulations of the periodic light curve resembling a rotating chemically peculiar stars with a period of 2.7 days (see Fig.\,\ref{LCshow}) shows possible pitfalls of standard processing of such type of data.

\subsection{Simulation of "TESS-like" data of a hypothetical CP star}

Chemically peculiar (CP) stars are rotating variables with extensive photometric spots on their surface. The observed light changes of such objects are strictly periodic with the period of star rotation and can be described by a low degree harmonic polynomial. For our example, we chose a two-wave curve represented by a third-degree harmonic polynomial, described by five opted parameters (\ref{polynom}) of the \emph{phase function}  $\vartheta$ \citep[for details see in][]{mik08}, $F(\vartheta,\textbf{a})$, $P=2.7$\,d, with a maximum in the phase $\varphi=0$.
\begin{eqnarray}
&\displaystyle\vartheta=\frac{t-M_0}{P}=E+\varphi,\quad M_0=1450+P\Delta \varphi,\quad E=\textrm{IP}(\vartheta),\quad \varphi=\textrm{FP}(\vartheta),\\
&F(\vartheta,\textbf{a})=a_1\,\cos(2\pi\vartheta)+a_2\,\cos(4\pi\vartheta)+a_3\, \cos(6\pi\vartheta)+ \label{polynom}\\
&+a_4\,[2\,\sin(2\pi\vartheta)-\sin(4\pi\vartheta)]
+a_5\,[3\,\sin(2\pi\vartheta)+6\,\sin(4\pi\vartheta)-5\,\sin(6\pi\vartheta)].\nonumber
\end{eqnarray}
where $t$ is the TESS BJD time of the observation ($t=\textit{BJD}-2\,457\,000$), $M_0$ is the TESS BJD moment of the initial light curve maximum, $\Delta \varphi$ is an optional \emph{initial phase} parameter allowing horizontal shift of the simulated light curve, $E$ is an integer epoch and $\varphi$ is a common phase. IP and FP means the integer part and  the fraction part of a number, $a_1=-5,\ a_2=-4.5,\ a_3=-0.5,\ a_4=-0.67,$ and $a_5=-0.17$\,mmag.

The simulated TESS light curve is represented by 17\,925 points with cadence of 2 minutes. The curve can, according to the assignment, show trends and Gaussian scatter (see Fig.\,\ref{LCshow}).

\begin{figure}
\centerline{\includegraphics[width=1\textwidth,clip=]{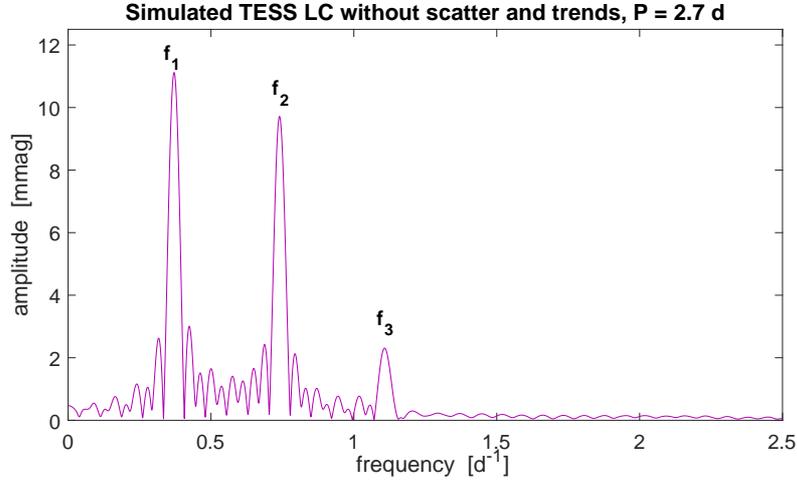}}
\caption{The amplitude periodogram of the simulated light curve without scatter and trends for $\Delta \varphi=0$.}
\label{period1}
\end{figure}

\section{The role of the initial phase $\Delta \varphi$}

The amplitude frequency spectrum of the simulated light curve displays as expected three dominant, equidistantly spaced peaks with central frequencies $f_1, f_2$, and $f_3$ (Fig.\,{\ref{period1}})), each carrying a period information: $1/f_1=2/f_2=3/f_3=P$. However, this is fulfilled only approximately. If we limit ourselves to two peaks, then for $\Delta\varphi = 0$ we get: $P_1 = 1/f_1 = 2.7070(10)$ d, $P_2=2/f_2=2.6977(8)$ d. The deviation from the baseline period $P=2.7$ d is thus evident and far exceeds the limits given by the uncertainty of the positioning of the frequency peaks. Why such a difference? The same results give other simple period finders.

The discrepancy would only disappear if the light curves were purely sinusoidal without higher harmonics. If they deviate from this ideal, the so-called 'periods' found are not real periods, but only parameters found by regression with an inadequate model that differ (sometimes fragrantly) from its pattern.

\begin{figure}
\centerline{\includegraphics[width=0.92\textwidth,clip=]{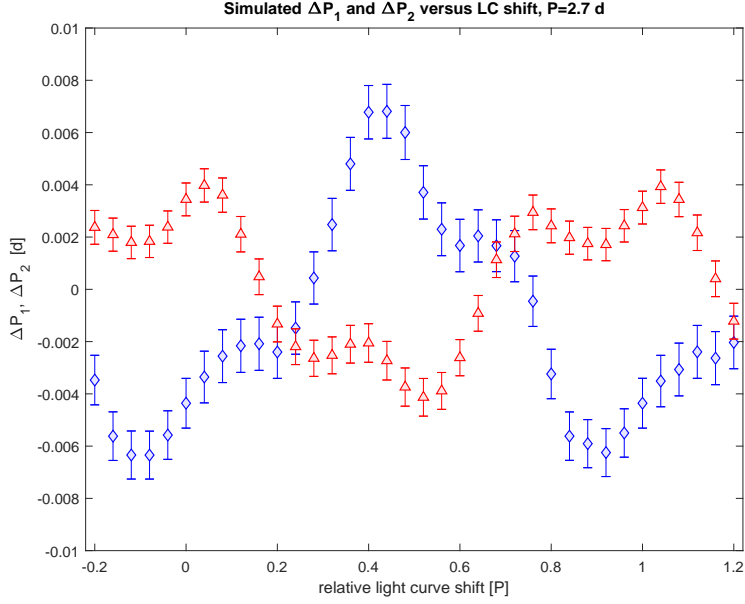}}
\caption{$\Delta P_1=P_1-P$ (blue diamonds) and $\Delta P_2=P_2-P$ (red triangles) differences as function of the light curve shift $\Delta \varphi$.}
\label{LC_shift}
\end{figure}

The following graph (Fig.\,\ref{LC_shift}) shows that the values of those 'periods' found in the periodograms are a complex periodic function of the initial phase $\Delta\varphi$, whose amplitude is simply shocking.

\subsection{Influence of light curve trends}

\begin{figure}
\centerline{\includegraphics[width=0.92\textwidth,clip=]{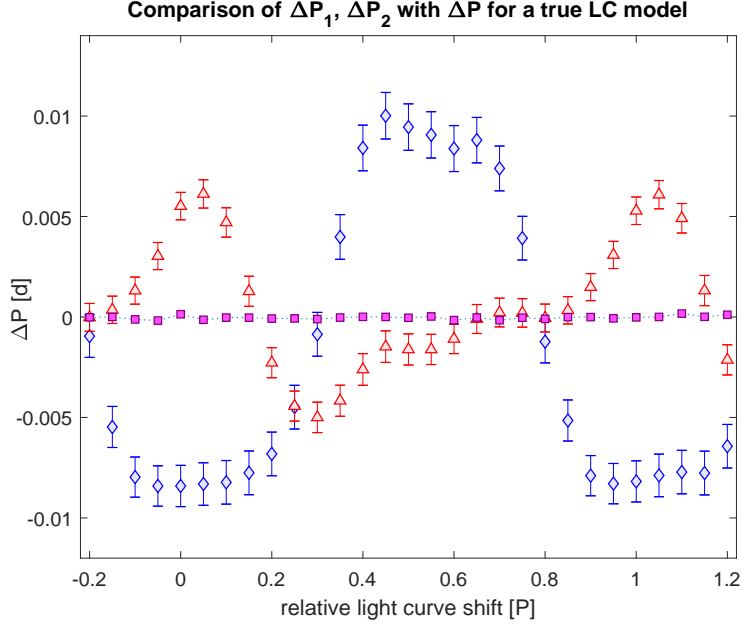}}
\caption{The comparison of $\Delta P_1$ (blue diamonds) and $\Delta P_2$ (red triangles) differences and true model ones (pink squares) for non-detrended increasing light curve as a function of the LC shift $\Delta \varphi$.}
\label{comparison}
\end{figure}

Both TESS and Kepler observations are strongly affected by aperiodic instrumental trends \citep{stefan18,mik19}. Neglecting them has a completely devastating effect on period analysis (Fig.\,\ref{comparison}). Appropriate detrending of the observed light curves is highly desirable, because only in this way can we utilize the unprecedented accuracy of satellite photometry.

\section{Modelling of light curves and trends}

When applying the Lomb-Scargle method, we have to adjust the observation data by subtracting their mean value from them. Subsequently, these data are fitted by the simplest possible model of the periodic light curve, i.e. a linear combination of a pair of harmonic functions $F=a_1\cos(2\pi f) + a_2 \sin(2\pi f)$, where the amplitude $A$, $A=\sqrt{a_1^2+a_2^2}$, is a function of the frequency $f$ and plotted on the $A(f)$ - amplitude periodogram, where the extremes of the plotted function are searched for and interpreted. It is even worse with estimating the uncertainty of determined frequencies with extreme amplitude values, because it is based on the apparatus of the least-squares method, but the conditions for its application are not met\footnote{Observed deviations of the functional value model does not have a normal distribution, consecutive functional values are not independent, etc.}.

It is apparent from the foregoing examples that the use of conventional low parametric models is not sufficient for the description of the high-precision real light curves provided by modern instruments, in particular the satellites. It is, therefore, essential to use more advanced models to simultaneously fit a phase curve with a harmonic polynomial of at least third degree, while modeling trends by dividing the light curve into segments and describing them with a polynomial of the appropriate degree \citep{mik19,stefan18}. Phase curve and trend models should be tailored to the actually observed light curves. The models should be functions of the parameters of the ephemeris of stellar periodicity, especially the time of the basic extremum, the mean period (or periods) and their time derivatives \citep{mik15,mik16}.

Modern tools such as chi-square approach, robust regression, bootstrapping, etc. can be used to find parameters (including period/periods) and estimate their uncertainties. The results of this rigorous procedure can be compared with the results obtained by standard trivial period analysis procedures in Fig.\,\ref {comparison}. No further comment is needed.

\section{Conclusions}
\begin{itemize}
  \item Short-term observational series corrupted by instrumental trends are not the easy observational material for accurate determination of periods.
  \item Using the Lomb-Scargle method we make serious errors both in the determination of the period and in the estimation of the uncertainty of this determination.
  \item In the periodic analysis of TESS data, the differences between the found period and the actual period are so huge that we are not able to correctly establish observations that are only a few months away.
\end{itemize}
A rigorous solution to the problem is to move to realistic models of light curves, including a true description of phase curve/curves as well as instrumental trends. Only then, we will fully use the information potential of short-term observation sets.

\acknowledgements
This work has been supported by grant GA\v{C}R 18-05665S and Operational Programme Research, Development and Education - Project "Postdoc@MUNI" (No. CZ.02.2.69/0.0/0.0/16\_027/0008360).

\bibliography{mik_caosp306}

\begin{thebibliography}{7}
\expandafter\ifx\csname natexlab\endcsname\relax\def\natexlab#1{#1}\fi

\bibitem[{{H{\"u}mmerich} {et~al.}(2018){H{\"u}mmerich}, {Mikul{\'a}{\v{s}}ek},
  {Paunzen}, {Bernhard}, {Jan{\'\i}k}, {Yakunin}, {Pribulla}, {Va{\v{n}}ko}, \&
  {Mat{\v{e}}chov{\'a}}}]{stefan18}
{H{\"u}mmerich}, S., {Mikul{\'a}{\v{s}}ek}, Z., {Paunzen}, E., {et~al.}, {The
  Kepler view of magnetic chemically peculiar stars}. 2018, {\it \aap}, {\bf
  619}, A98, DOI: 10.1051/0004-6361/201832938

\bibitem[{{Mikul{\'a}{\v{s}}ek}(2015)}]{mik15}
{Mikul{\'a}{\v{s}}ek}, Z., {Phenomenological modelling of eclipsing system
  light curves}. 2015, {\it \aap}, {\bf 584}, A8, DOI:
  10.1051/0004-6361/201425244

\bibitem[{{Mikul{\'a}{\v{s}}ek}(2016)}]{mik16}
{Mikul{\'a}{\v{s}}ek}, Z., {Monitoring of rotational period variations in
  magnetic chemically peculiar stars}. 2016, {\it Contributions of the
  Astronomical Observatory Skalnate Pleso}, {\bf 46}, 95

\bibitem[{{Mikul{\'a}{\v{s}}ek} {et~al.}(2008){Mikul{\'a}{\v{s}}ek},
  {Krti{\v{c}}ka}, {Henry}, {Zverko}, {{\v{Z}}i{\v{z}}{\r{a}}ovsk{\'y}},
  {Bohlender}, {Romanyuk}, {Jan{\'\i}k}, {Bo{\v{z}}i{\'c}},
  {Kor{\v{c}}{\'a}kov{\'a}}, {Zejda}, {Iliev}, {{\v{S}}koda}, {{\v{S}}lechta},
  {Gr{\'a}f}, {Netolick{\'y}}, \& {Ceniga}}]{mik08}
{Mikul{\'a}{\v{s}}ek}, Z., {Krti{\v{c}}ka}, J., {Henry}, G.~W., {et~al.}, {The
  extremely rapid rotational braking of the magnetic helium-strong star HD
  37776}. 2008, {\it \aap}, {\bf 485}, 585, DOI: 10.1051/0004-6361:20077794

\bibitem[{{Mikul{\'a}{\v{s}}ek} {et~al.}(2015){Mikul{\'a}{\v{s}}ek}, {Paunzen},
  {Netopil}, \& {Zejda}}]{mikpau15}
{Mikul{\'a}{\v{s}}ek}, Z., {Paunzen}, E., {Netopil}, M., \& {Zejda}, M., {New
  Tools for Finding and Testing Weak Periodic Variability}. 2015, in
  Astronomical Society of the Pacific Conference Series, Vol. {\bf  494}, {\it
  Physics and Evolution of Magnetic and Related Stars}, ed. Y.~Y. {Balega},
  I.~I. {Romanyuk}, \& D.~O. {Kudryavtsev}, 320

\bibitem[{{Mikul{\'a}{\v{s}}ek} {et~al.}(2019){Mikul{\'a}{\v{s}}ek}, {Zejda},
  {H{\"u}mmerich}, {Krti{\v{c}}ka}, {Bernhard}, {Paunzen}, {Skarka}, {de
  Villiers}, {Jagelka}, \& {Baki{\textcommabelow s}}}]{mik19}
{Mikul{\'a}{\v{s}}ek}, Z., {Zejda}, M., {H{\"u}mmerich}, S., {et~al.},
  {Monitoring Period Variations of Variable Stars using Precise Photometric
  Surveys}. 2019, in IAU Symposium, Vol. {\bf  339}, {\it Southern Horizons in
  Time-Domain Astronomy}, ed. R.~E. {Griffin}, 110--113

\bibitem[{{Press} \& {Rybicki}(1989)}]{press89}
{Press}, W.~H. \& {Rybicki}, G.~B., {Fast Algorithm for Spectral Analysis of
  Unevenly Sampled Data}. 1989, {\it \apj}, {\bf 338}, 277, DOI: 10.1086/167197

\end{thebibliography}
\end{document}